
\magnification1200
\rightline{KCL-MTH-06-07, UB-ECM-PF 06/19, TIT/HEP-553,
Toho-CP-0681} 
\rightline{hep-th/0607057}

\vskip 1.5cm \centerline {\bf    The construction of  brane and
superbrane actions  using non-linear realisations} \vskip 1cm
\centerline{Joaquim Gomis} \centerline{Departament ECM, Facultat
F\'{\i}sica} \centerline{Universitat Barcelona, Diagonal 647}
\centerline{E-08028 Barcelona, Spain}
\centerline{Department of Physics} \centerline{Tokyo Institute of
Technology} \centerline{Meguro, Tokyo 152-8551, Japan} \centerline{}
\centerline{Kiyoshi Kamimura} \centerline{Department of
Physics}\centerline{Toho University, Funabashi 274-8510, Japan}
\centerline{} \centerline {and} \centerline{}\centerline{Peter 
West} \centerline{Department of Mathematics} \centerline{King's
College, London WC2R 2LS, UK}

\vskip 0.5cm
\vskip .2cm \noindent We derive new  actions for the bosonic
p-brane, super p-brane and the p-brane moving in AdS (dS)
space-times using the theory of non-linear realisation without
requiring the adoption of any constraints or using superfields. The
Goldstone boson associated with the breaking of Lorentz
transformations becomes a dynamical field whose equation of motion
relates it algebraically to the remaining  Goldstone fields.

\vskip .5cm

\vfill \eject {\bf 1 Introduction}

Non-linear realisations played a crucial role to understand the
importance  of symmetries in particle physics. In particular
it has been  developed in constructing lagrangians with
chiral symmetries [1]. 
The classic  papers of Callan, Coleman, Wess and
Zumino [2] set out the general theory of non-linear realisations
that applied to internal symmetries. In this case one takes  a group
$G$ with  a subgroup $H$ and considers  the coset  elements of $G/H$
which are taken to depend on space-time, which is otherwise
unconnected with the group $G$. Transformations of the chosen coset
representatives are then preserved by compensating $H$
transformations. A slightly more general viewpoint than that
considered contained in the original papers of [2] is to take a
group $G$ and work with group elements which depend on space-time,
but take the theory to be invariant under the two independent
transformations. $g(x^\mu)\to g_0 g(x^\mu) ,\ \ {\rm and }\ \
g(x^\mu)\to g(x^\mu) h(x^\mu)$ where $g_0$ is a rigid transformation
which belongs to $G$ and $ h(x^\mu)$ is a local transformation that
belongs to $H$ [3,4]. In this way of proceeding one can refer to a
non-linear realisation of a group $G$ with local subgroup $H$. The
dynamics is usually constructed  from the Cartan forms which are
clearly inert under rigid $g_0$ transformations and only transform
under local $H$ transformations. Clearly, the Cartan forms belong to
the Lie algebra of $G$, but when decomposed to representations of
the sub-algebra $H$ they contain the adjoint as well as other
representations of $H$. While the adjoint representation of $H$
contained in the Cartan form $g^{-1}dg$ transforms inhomogeneously,
all the remaining representations transform linearly and can be used
to construct an action. If one is interested in the lowest energy
contributions then one must construct an action with  two space-time
derivatives, that is an action that is only  bilinear in the Cartan
forms.  Since the latter only transforms under local $H$
transformations the construction of the most general such action is
a problem in finding invariants under the subgroup $H$ of the
representations that arise in the Cartan form. Although well known
to experts in non-linear realisations, it is not widely appreciated
at large that the  result is  unique up to a few possible constants.
Indeed one finds one constant for every quadratic numerically
invariant $H$ tensor  that occurs for  the representations, not
including the adjoint,  that occurs in  the decomposition of the
Cartan form. For a sufficiently large subgroup $H$,  there is only
one such numerically invariant tensor and the result is completely
fixed.
\par
Branes can be viewed as the motion of solitonic objects that occur
in field theories and from this perspective their motion can be seen
as a non-linear realisation corresponding to the symmetries that are
spontaneously broken by the soliton. Clearly, the embedding
coordinates transverse to the brane are the Goldstone bosons for the
broken translations and in a supersymmetric theory part of the
fermions correspond to the breaking of the supersymmetries by the
solitons. Many of the early discussions of deriving the dynamics of
branes from the theory of non-linear realisations  concerned
particular supersymmetric branes in space-times in  low dimensions
compared to that currently of most interest.  In the first such
papers [5]  supersymmetric branes in four and six dimensional
space-times were considered, but  Goldstone bosons   were introduced
for  only  the breaking of translations, central charges  and
supersymmetries. However, in another early  paper [6]   the bosonic  point
particle  in two dimensions was considered and  Goldstone bosons
corresponding to the broken translations, Lorentz transformations
were introduced and an  action was given  that depended on both of
these Goldstone fields. It turned out that   the equation of motion
of the Goldstone field for Lorentz transformations expressed it in
terms of the Goldstone bosons for the broken translations. Thus this
equation of motion  was an example of the inverse Higgs effect [7]
which uses covariant constraints to eliminate algebraically 
Goldstone fields in terms of the others. These considerations were
generalised to the non-relativistic particle and the massive
superparticle in two dimensions where in both cases it was found
that the actions can be interpreted as Wess-Zumino terms [8].
Later studies [9] of particular supersymmetric branes in several
different dimensions also included Goldstone bosons corresponding to
the breaking of generators of the Lorentz and some internal
symmetries and the inverse Higgs mechanism was used to eliminate these
Goldstone bosons.  In all the papers mentioned above the unbroken
translations and supersymmetry generators occurred in the group element
associated with coordinates, which turned out to parameterise the brane
world, and the  Goldstone fields were taken to depend on these
coordinates.  Actions for p-branes using  Lorentz harmonics have 
also been constructed from the geometrical viewpoint of the theory of 
embeddings [10]. 
\par
The general procedure for using the theory of the non-linear
realisation to construct the dynamics of branes is to take a  group
that has  a semi-direct product structure.   In particular,   we
consider   a group $K$, with a specified subgroup $H$, together with
a representation $L$ which form a semi-direct product group denoted
$G=K\otimes_s L$.  We then further sub-divide the generators in the
representation $L$ of $K$ into two sets $L_1$ and $L_2$ which are
representations of  $H$,  a sub-algebra of $K$.  The non-linear
realisation of interest is that based on the group $G$ with local
subgroup $H$. We may think of the algebra $K$ as an automorphism
algebra of the algebra $L$ which is not always a commuting algebra.
\par
The dynamics of the simple bosonic p-brane from the view point of
non-linear realisations was only considered  relatively recently
[11]. For this case one takes $K=SO(1,D-1)$ and $L$ are the
translation generators in the vector representation of
$K=SO(1,D-1)$. We then take $L_1$ to contain   $p+1$ of the
translation generators and $H=SO(1,p)\otimes SO(D-p-1)$.  The
generators of $L_1$ being a vector under the first factor and a
scalar under the second factor of $H$.  The sets  $L_1$ and $L_2$
contain the unbroken and broken translations and  $H$ is the
group of unbroken Lorentz rotations. The Goldstone bosons of the Lorentz
transformations played an important role and it was found they could
be eliminated by adopting a covariant constraint (inverse Higgs)  in
terms of the Goldstone bosons for broken translations.  Reference [11]
also introduced  fields associated with the unbroken translations
which appeared in the group element and these in
common with all the Goldstone fields were taken to depend on external
coordinates that turned out to parameterise the brane world volume. One
could then construct an action, after the elimination of the Goldstone
bosons for broken Lorentz transformations by the inverse Higgs  mechanism,
which was invariant under the transformations of the non-linear
realisation and in addition demand that the theory was  reparameterisation
invariant.  In the supersymmetric cases
studied  in [11] the Goldstone fields were also taken to depend on
Grassman  odd coordinates and so were  superfields.
\par
It has been advocated [11] that to find the dynamics of certain
supersymmetric branes using the theory of non-linear realisations one
should take for $K$ the most  general automorphism algebra of $L$ rather
than just the Lorentz  and internal symmetry algebras previously
considered. In particular, it was found [11,12] that the field
strengths of the gauge fields that occur in D-branes are the
Goldstone bosons associated with certain broken generators of  such
an enlarged automorphism algebra and that the gauge fields
themselves are the Goldstone fields associated with certain central
charges in the supersymmetry algebra. In fact this is an example of
the general case;   the Goldstone bosons associated with the broken
generators of the automorphism algebra are often space-time
derivatives of the Goldstone bosons associated with the broken
generators of the algebra $L$ [12].
\par
Non-relativistic branes for Galilei [13] and Newton Hooke [14]
groups have been constructed using non-linear realizations along the
lines of the  previous paragraph. The corresponding actions are
Wess-Zumino terms of the  previous groups.
\par
In this paper we return to the construction of  the actions for
bosonic and superbranes from the theory of non-linear realisations.
We will introduce fields  for all the translations, the broken
Lorentz symmetries and in the supersymmetric case for the  all
supercharges.  All these fields will then be taken to depend on a
set of parameters $\xi$ that will turn out to parameterise the brane
world volume. We will give a new action  constructed from all of
these Goldstone fields which is invariant under the transformations
of the non-linear realisation and reparameterisations in the
parameters $\xi$.  The equations of motion of the Goldstone fields
corresponding to the broken Lorentz generators express these fields
in terms of the remaining Goldstone fields. Using these equations of
motion to algebraically eliminate the Goldstone bosons associated
with the broken Lorentz transformations   in the action one finds
the standard results for the dynamics of  bosonic and super
p-branes. All previous discussions of the dynamics of
supersymmetric branes using non-linear realisations have used
superfields. However, in this paper we will take the Goldstone
fields to depend on a set of Grassman even parameters alone and so
we will not  use superfields. That we can proceed in this way is a
consequence of the  fact that in a non-linear realisation all the
fields  that appear in the superbrane are Goldstone fields  and so
are automatically introduced  into the group element  by the
corresponding generator in the algebra.   One of the motivations for
this work is to give a systematic procedure for finding the dynamics
of the simplest relativistic branes  of interest in the hope that
this will provide  useful lessons that can be used for more
complicated systems.

\medskip
{\bf 2 p-Brane in Flat Background}
\medskip
We now  construct the non-linear realisation that leads to dynamics
of  the bosonic p-brane moving in a $D$-dimensional Minkowski
space-time and so consider the algebra of translations and Lorentz
rotations,  $G=ISO(1,D-1)$, whose generators obey the commutators
$$
[J_{\underline a\underline b}, J_{\underline c\underline d}]=
-i\eta_{\underline b\underline c}J_{\underline a\underline d}
+i\eta_{\underline a\underline c}J_{\underline b\underline d}
+i\eta_{\underline b\underline d}J_{\underline a\underline c}
-i\eta_{\underline a\underline d}J_{\underline b\underline c}
\eqno(2.1)$$
$$
[J_{\underline a\underline b}, P_{\underline c}]=-i \eta_{\underline
b\underline c}P_{\underline a}+ i\eta_{\underline a\underline
c}P_{\underline b}. \eqno(2.2)$$ We take as our local sub-algebra
$H=SO(1,p)\otimes SO(D-p-1)$ which contains the generators $ J_{ab},
J_{a'b'}$ which are  the unbroken Lorentz rotations. We are using
the notation that an underlined index, i.e. $ \underline a$ goes
over all possible values from $0$ to $D-1$ while the unprimed
indices $a$ take the values  $a=0,\ldots , p$ and primed indices
$a'$ take the values $p+1,\ldots , D-1$. The latter are the indices
which are longitudinal and transverse to the brane respectively.
\par
The group elements  are taken to depend on the parameters $\xi^i, \
i=0,\ldots , p$. The non-linear realisation is by definition just
the theory  invariant under the  transformations
$$
g(\xi)\to g_0 g(\xi) ,\ \ {\rm and }\ \  g(\xi)\to g(\xi) h(\xi)
\eqno(2.3)$$ where $g_0$ is a rigid transformation which belongs to
$ISO(1,D-1)$ and $ h(\xi)$ is a local transformation that belongs to
$H=SO(1,p)\otimes SO(D-p-1)$.
\par
We can take the group element to be given by
$$
g=e^{ix^{\underline a} P_{\underline a}} e^{i\phi_{a}{}^{b'}
J^{a}{}_{b'}}\; , \eqno(2.4)$$ where  $x^{\underline a}$ and
$\phi_{a}{}^{b'}$ depend on the parameters $\xi^i$.  We note that
this is not the most general group  element as we have used the
local Lorentz transformations to set part of the group element to
one.
\par
It is convenient to take use an explicit form of the $SO(1,D-1)$
Lorentz transformations that is particularly suited to the
decomposition into $SO(1,p)\otimes SO(D-p-1)$.  Let us write the
$SO(1,D-1)$ Lorentz transformation as
$$
e^{-i\lambda\cdot J} e^{-i\phi_{ b}{}^{ c'}  J^{ b}{}_{ c'}} P_{ a}
e^{i\phi_{ b}{}^{ c'}  J^{ b}{}_{ c'}}e^{i\lambda \cdot J}=
\Phi_{\underline a}{}^{\underline c} L_{\underline c}{}^{\underline
b} P_{\underline b}
\eqno(2.5)$$ where $\lambda \cdot J={1\over 2}(
\lambda _{ b}{}^{ c}  J^{ b}{}_{
c}+\lambda_{ b'}{}^{ c'} J^{ b'}{}_{ c'})$ is a general
$SO(1,p)\otimes SO(D-p-1)$ generator and $L_{\underline
a}{}^{\underline c}$ is the corresponding finite transformation on
$P_{\underline a}$. We  may express $\Phi$ in the form
$$
\Phi= \left(\matrix{I_1& \varphi \cr -\varphi^T &
I_2\cr}\right)\left(\matrix{B_1& 0\cr 0 &
B_2\cr}\right)=\left(\matrix{B_1& \varphi B_2 \cr -\varphi^TB_1 &
B_2\cr}\right) \eqno(2.6)$$ where
$B_1=(I_1+\varphi\varphi^T)^{-{1\over 2}}$ and
$B_2=(I_2+\varphi^T\varphi)^{-{1\over 2}}$. In this equation
$\varphi$ carries the indices $\varphi_a{}^{b'}$ and so the matrices
$B_1$   and $B_2$   have the indices $(B_1)_{a}{}^{b}$ and
$(B_2)_{a'}{}^{b'}$ respectively and similarly for the unit matrices
$I_1$ and  $I_2$.
Clearly, the  Lorentz transformations where $\lambda _{ b}{}^{ c}
=0=\lambda_{b'}{}^{ c'}$ are just given by $\Phi$ and
$\varphi_a{}^{b'}$ parameterise the coset $SO(1,D-1)/ SO(1,p)\otimes
SO(D-p-1)$. The inverse Lorentz transformation is given by
$$
\Phi^{-1}=\left(\matrix{B_1& -B_1\varphi  \cr B_2\varphi^T &
B_2\cr}\right)\;. \eqno(2.7)$$
\par
The dynamics is  built out of the Cartan forms
$$
{\cal V}= -i g^{-1}dg={\cal V}_i d\xi^i
$$
where ${\cal V}_i$ is given by
$$
{\cal V}_i= e_i{}^a P_{a}+e_i{}^{a'} P_{a'}+{1\over 2} w_{i
\underline a}{}^{\underline b}   J^{\underline a}{}_{\underline b}
\eqno(2.8)$$ which are inert under rigid $G$ transformations and
transform under local $SO(1,p)\otimes SO(D-p-1)$  transformations as
$$
{\cal V}_i\to h^{-1}{\cal V}_i h-i h^{-1}{\partial}_i h\;.
\eqno(2.9)$$ Of course if one is working with a group element part
of  which   is fixed by a local $H$ transformation, as indeed is the
case here, then a rigid $G$ transformation involves a compensating
$H$ transformation. However, when constructing invariants this can
be viewed as just another local $H$ transformation. The Cartan forms
are given by
$$
e_i{}^a=\partial_i x^{\underline c}\Phi_{\underline c}{}^{ a}=
(\partial_i x^c-\partial_ix^{b'}\varphi^T_{b'}{}^{c})(B_1)_c{}^a,\
$$
$$
e_i{}^{a'}=\partial_i x^{\underline c}\Phi_{\underline c}{}^{a '}=
(\partial_i x^c \varphi_{c}{}^{b'}+\partial_i
x^{b'})(B_2)_{b'}{}^{a'},\  w_{i\underline a}{}^{\underline b}
=(\Phi^{-1})_{\underline a}{}^{\underline c}\partial_i (\Phi)_
{\underline c}{}^{\underline b}\;. \eqno(2.10)$$ The Cartan forms
transform under the local $SO(1,p)\otimes SO(D-p-1)$ as their
indices suggest; for example if $h= e^{i{1\over 2} \beta_{a}{}^{b}
J^{a}{}_{b}}$ then
$$
\delta e_i{}^a=e_i{}^c \beta_{c}{}^{a} ,\ \delta e_i{}^{a'}=0,\;
$$
$$
\delta w_{ia}{}^{ b}=
\partial_i\beta_{ a}{}^{ b}+
w_{i a}{}^{ c}\beta_{c}{}^{b}-\beta_{a}{}^{c}w_{ic}{}^{b},\; \delta
w_{i b'}{}^{\underline a}= w_{i\underline b'}{}^{
c}\beta_{c}{}^{a}\;.
    \eqno(2.11)
$$
We can interpret $e_i{}^a$ as a vielbein on the brane world volume.
\par
We require an  action which is invariant under the transformations
of the non-linear realisation of equation (2.4)  and, imposed as an
additional requirement,  is also invariant under diffeomorphisms in
$\xi^i$. We therefore take as our action
$$
\int d^{p+1}\xi \det e_i{}^a \;. \eqno(2.12)$$
\par
The   action of equation (2.12) is a functional of
$\varphi_a{}^{b'}$ and $x^{\underline b}$. Using the expression of
equation (2.10) we find that we  may write the action as
$$
\int d^{p+1}\xi \det ({\partial x^c\over \partial \xi^i}) \det
(I_1-\partial x\varphi ^T) \det (I_1+\varphi\varphi^T)^{-{1\over 2}}
\eqno(2.13)$$ where $\partial x$ is the matrix $(\partial
x)_{a}{}^{b'}={\partial x^{b'}\over \partial x^a}$.  Varying with
respect to $\varphi_a{}^{b'}$  we find the equation of motion
$$
(I_1+\varphi\varphi^T)^{-1}\varphi=-(I_1- \partial x\varphi^T)^{-1}
\partial x  \;.
\eqno(2.14)$$ In deriving this equation we have used the fact that
$(I_1+\varphi\varphi^T)$ is a symmetric matrix. Multiplying by
$\varphi^T$ we find that
$$
(I_1+\varphi\varphi^T)^{-1}=(I_1- \partial x\varphi^T)^{-1} \;.
\eqno(2.15)$$ Substituting this   result in equation (2.14) we
conclude that the unique solution is
$$
\varphi=-\partial x, \ \ {\rm or } \ \ \varphi_a{}^{b'}=-{\partial
x^ {b'} \over \partial x^a} \;.  \eqno(2.16)$$ This latter equation
is just   solution to the condition
$$
e_i{}^{a'}=0 \;.  \eqno(2.17)$$
\par

We could have adopted this condition as an inverse Higgs condition
from the outset as it is an invariant condition under the
transformations of the non-linear realisation. Using (2.16) we find
that
$$
e_i{}^a=(\partial_i x^b)((I_1+\partial x(\partial x)^T)^{{1\over
2}})_b{}^a \eqno(2.18)$$ and the action (2.12) becomes the covariant
Nambu-Goto action in flat space
$$
A=
-T\int\;d^{p+1}\xi\;\sqrt{-\det G_{ij}}
 \eqno{(2.19)}$$ where
$G_{ij}$ is the induced metric
$$
G_{ij}= {\partial_i}x^{\underline a}{\partial_j}x^{\underline b}
\eta_{{\underline a}{\underline b}} \;. \eqno{(2.21)}$$

{The Goldstone bosons for the Lorentz generators are the Lorentz  
harmonics used in reference [10]  to construct actions for  p-branes. 
However, these authors also introduce a vielbein on the  brane world sheet 
and do not have an explicit parameterisation for  their Lorentz harmonics, 
but rather work with constrained fields. 
The  action of equation (2.12) may well be related to the actions used in  
reference [10] provided one eliminates some fields using their  
algebraic equations of motion and introduces the parameterisation  
used here for Lorentz transformations. }

\vskip 3mm

{\bf 3. Relativistic Brane in  AdS$_D$ and dS$_D$ : }

\medskip
We can apply the method of the non-linear realization to find the
dynamics of a p-brane in the AdS$_D$ ( dS$_D$) space-time. As such
we take $G$ to be the AdS$_D$ (dS$_D$) algebra which consists of the
generators $J_ {\underline a\underline b}$, that  obey the
commutator of equation (2.1) as well as  the generators
$P_{\underline a}$ that  can be viewed as the translations, although
they  are not commuting
$$
\left[P_{\underline a},~P_{\underline b}~\right]=\mp
i{{1}\over{R^2}} \;J_{\underline a\underline b}  \;.
\eqno{(3.1)}$$ In this last equation $-$ refers to AdS$_D$ and $+$
to dS$_D$ and $R $ is the radius of  the AdS$_D$ ( dS$_D$) space. We
take the local sub-algebra  to be $H=SO(1,p)\otimes SO(D-p-1)$ which
is the same as in the case of flat space. The group element is
subject to the transformation of equation (2.3).  Using the local
symmetry, the group elements can be chosen to be of the form
$$
g=g_0\;U,\qquad g_0:=e^{ix^{\underline a} P_{\underline a}},\qquad
U:=e^{i\phi_{a}{}^{b'}  J^{a}{}_{b'}}\eqno(3.2).
$$
where the fields $x^{\underline a} $ and $\phi_{a}{}^{b'}$ depend on
the parameters $\xi^i,( i=0,1.\ldots ,p)$. The Cartan  one forms are
given by
$$
{\cal V}=U^{-1}{\cal V}_0 U-iU^{-1}d U
=e^{\underline a}P_{\underline a}+{1\over2} w^{{\underline
a}{\underline b}}J_{{\underline a}{\underline b}} \;. \eqno{(3.3)}$$
The expression ${\cal V}_0$ that occurs in  the first term is given
by
$$
{\cal V}_0=-ig_0^{-1}dg_0=\tilde e^{\underline a}P_{\underline a}+{1
\over2} \tilde w^{{\underline a}{\underline b}}J_{{\underline
a}{\underline b}}, \eqno{(3.4)}$$ where $\tilde e^{\underline a}$
and $\tilde w^{\underline ab}$ are AdS$_{D}$(dS$_D$) vielbein and
spin connection which satisfy the Maurer Cartan equations of the
AdS$_{D}$(dS$_D$) algebra
$$
d\tilde e^{\underline a}+\tilde w^{{\underline{ab}}}\tilde e_
{\underline b}=0, \qquad d\tilde w^{{\underline{ab}}}+{\tilde
w^{\underline a}}{}_{\underline c} \tilde w^{\underline{cb}}\pm
{{1}\over{R^2}}\tilde e^{\underline a}\tilde e^{\underline b}=0 \;.
\eqno{(3.5)}$$ The Cartan forms of equation (3.3) are inert under
the rigid transformations, but transform under the local
transformations as in equation (2.9). Using equation (2.5) we find
that
$$
U^{-1}P_{\underline a}U= {\Phi_{\underline a}}^{\underline b}
P_{\underline b}, \eqno{(3.6)}$$ and so we conclude that
$$
e^{\underline a}=\tilde e^{\underline b}{\Phi_{\underline b}}^
{\underline a} \eqno{(3.7)}$$ where ${\Phi_{\underline
b}}^{\underline a}$ is given in equation (2.6). The  expression for
$e^{\underline a}$ is obtained from those of the Poincare case by
just substituting $\partial_i x^{\underline a} $ by ${\tilde
e_i}{}^{\underline b}$ and hence the way $\varphi_a{}^{b'}$ occurs
is similar.

The action should be invariant under the transformations of the
non-linear realisation, that is the local and rigid transformations
of equation (2.3). However, we also demand that it be invariant
under diffeomorphism in  the parameters $\xi^i$. The  1-forms $e^a$
are inert under the rigid transformations and transform as vectors
under the local $SO(1,p)$ transformations. Hence,  the action is
given by
$$
A=-T \int \Omega_{p+1}=-T\int\;d^{p+1}\xi \det({e_i}^a)
\eqno{(3.8)}$$ where
$$
\Omega_{p+1}=-{1\over{(p+1)!}}\epsilon_{a_0...a_p}e^{a_0}\wedge ...
\wedge e^{a_p} , \eqno{(3.9)}$$ we take
$\epsilon^{01...p}=+1=-\epsilon_{01...p}.$

The action  depends on the non-dynamical Goldstone field
${\varphi_a}^ {b'}$ in a similar way to that for the case of  flat
space. Following similar  steps as in the previous section we
conclude that the equations of motion for ${\varphi_a}^{b'}$ leads
to the condition
$$
{\varphi_a}^{b'}=-{(\tilde e^{-1})_a}^i{\tilde e_i}{}^{b'}  \;.
\eqno{(3.10)}$$ This is equivalent to imposing  the inverse Higgs
conditions
$$
{e_i}^{a'}=0  \;.  \eqno{(3.11)}$$
\par
Substituting the expression for ${\varphi_a}^{b'}$ of equation
(3.10) into the action we get the known  generalisation of the
Nambu-Goto action to a bosonic brane moving in AdS$_D$(S$_D$)
space-time
$$
A=
-T\int\;d^{p+1}\xi\;\sqrt{-\det G_{ij}}
\eqno{(3.12)}$$ where $G_{ij}$ is the induced metric
$$
G_{ij}={{\tilde e_i}}{}^{\underline a}{{\tilde e_j}}{}^{\underline
b} \eta_{{\underline a}{\underline b}}:= {\partial_i}x^{\underline
m}{\partial_j}x^{\underline n} g_{{\underline m}{\underline n}}(x),
\eqno{(3.13)}$$ where ${\tilde e_i}{}^{\underline
a}=\partial_ix^{\underline m} {\tilde e_{\underline
m}}{}^{\underline a}$  and $g_{{\underline m}{\underline n}}(x)=
\tilde e_{\underline m}{}^ {\underline a}
\tilde e_{\underline n} {}^{\underline b}\eta_{\underline
a\underline b}$  is the
metric on AdS$_D$ (dS$_D$)   space-time.
\medskip
\vskip 3mm

{\bf 4. Super p-Branes }
\medskip
We now wish to  use the theory of non-linear realisations to find
the dynamics for the super p-brane in flat space. To this end we
take the algebra  $G$ to be the super Poincare algebra whose
generators obey the commutators of equations (2.1) and (2.2) as well
as the relations
$$
\left[Q_{\underline \alpha},~J_{{\underline{a b}}}\right] = -
{{i}\over{2}}  (\Gamma_{\underline{ab}})_{\underline \alpha}{}^
{\underline \beta} Q_{\underline \beta}, \quad \left[Q_{\underline
\alpha},~P_{\underline{a}}\right] = 0, \quad \{ Q_{\underline
\alpha},Q_{\underline \beta}\} = 2(\Gamma^ {\underline
a}C^{-1})_{\underline \alpha\underline \beta} \;P_{\underline {a}},
\eqno(4.1)$$ where $C$ is the charge conjugation matrix.

We take as local sub-algebra $H=SO(1,p)\otimes SO(D-p-1)$ which is
the same algebra as for bosonic branes. The group elements can be
chosen to be of the form
$$
g=g_0\;U,\qquad g_0:=e^{ix^{\underline a} P_{\underline
a}}\;e^{\bar\theta^{\underline \alpha}Q_{\underline \alpha} },\qquad
U:=e^{i\phi_{a}{}^{b'} J^{a}{}_{b'}} \;,
\eqno(4.2)$$
and  $\bar\theta^{\underline\alpha}=\theta_{\underline
\beta}C^{\underline \beta\underline\alpha}$.  Note that we take
the group element $g$, and so the fields $x^{\underline a} $,
$\theta_{\underline \alpha}$ and
$\phi_{a}{}^{b'}  $,  to depend on the Grassman even parameters
$\xi^i,i=0,1,\ldots ,p$. This differs from the usual  treatment
where one usually takes the fields to depend on the parameters
$\xi^i $ as well as a set of Grassman odd variables
$\theta_{\underline \alpha}$. The group element then contains superfields. In a
usual supersymmetric theory the component fields of the theory are
contained as the coefficients of the superfields when Taylor
expanded in the $\theta$. However, in a non-linear realisation all
the fields that arise correspond to the breaking of some generator
of the algebra and so will occur as the lowest component of the
superfield that is associated with that generator. Hence, using
superfields in the theory of non-linear realisations one finds a
considerable redundancy as all the component fields that occur  as
higher components of the superfields will also occur at the  bottom
of a superfield. This redundancy will imply the existence of   a
number of inverse Higgs conditions which will relate the
superfields. Taking the fields to depend on just the $ \xi^i$, as we
do here,  we will avoid this redundancy and allow us to  work with
the minimal of formalism.

The Cartan   forms are given by
$$
{\cal V}=U^{-1}{\cal V}_0 U-iU^{-1}d U=
e^{\underline a}P_{\underline a}-i\bar e^{\underline \alpha}
Q_{\underline \alpha}+{1\over2} w^{{\underline a}{\underline
b}}J_{{\underline a}{\underline b}}\;.
\eqno{(4.3)}$$
The expression
${\cal V}_0$ which appears in the first term is given by
$$
{\cal V}_0=-ig_0^{-1}dg_0=\pi^{\underline a}P_{\underline
a}-i\bar\pi^{\underline \alpha} Q_{\underline \alpha}
\eqno{(4.4)}$$ where $\pi^{\underline a}$, $\bar\pi^{\underline \alpha}$
are supervielbeins of Poincare superspace which are given by
$$
\pi^{\underline
a}=dx^{{\underline{a}}}+i\bar\theta\Gamma^{\underline a}
d\theta,\qquad\qquad \bar\pi^{\underline \alpha}=d\bar\theta^{\underline
\alpha} \;.  \eqno{(4.5)}$$
We recall that the spin connection of Poincare
superspace vanishes.
Using (2.5) and the  analogous equation for the spinor
representation ${\tilde\Phi_{\beta}}{}^{\alpha}$ of the boost $U$
$$
U^{-1}P_{\underline a}U= {\Phi_{\underline a}}^{\underline b}
P_{\underline b},\qquad U^{-1}Q_{\alpha}U=
{\tilde\Phi_{\alpha}}{}^{\beta}Q_{\beta}, \eqno{(4.6)}$$
$$
e^{\underline a}=\pi^{\underline b}{\Phi_{\underline b}}^{\underline
a}, \qquad\qquad \bar e^{\underline \alpha}=\bar \pi^{\underline
\beta}{\tilde\Phi_{\underline \beta}}{}^{\underline \alpha}.
\eqno{(4.7)}$$ where ${\Phi_{\underline b}}^{\underline a}$ is given
in equation (2.6)  and $\tilde \Phi_{\underline
\alpha}{}^{\underline \beta}=\left(\exp ({i\over 2} \phi\cdot
J)\right)_{\underline \alpha}{}^{\underline \beta}= \left(\exp (i
\phi_a{}^{b'} J^a{}_{b'})\right)_{\underline \alpha}{}^{\underline
\beta}= \left(\exp ( {1\over 2}\phi_a{}^{b'}
\Gamma^a{}\Gamma_{b'})\right)_{\underline \alpha}{}^{\underline
\beta}$. We note that the expression for $e^{\underline a}$ is the
same as that of equation (2.10) except that $dx^{\underline a}$ is
replaced by $\pi^{\underline a}$.

The action for the super p-brane must be invariant under the rigid
and local transformation of equation of the non-linearly realised
theory, however, we will also demand that it is invariant under
diffeomorphisms in the parameters $\xi^i$. The quantity $e^a$
transforms as a vector under $SO(1,p)$ transformations and so the
action
$$
A^{NG}=-T\int {\cal L}^{NG},\qquad
{\cal L}^{NG}=-{1\over{(p+1)!}}\epsilon_{a_0,...,a_p}e^{a_0}\wedge
.... \wedge e^{a_p}, \eqno{(4.8)}$$ is invariant.

However, there is another invariant action due to the existence of a
non-trivial  $p+2$ form in  the Chevalley Eilenberg cohomology group
[15]. The  Chevalley Eilenberg cohomology considers the group
manifold and the space of closed forms that are invariant under the
left action of the group.  We take two closed forms $h$ and $h'$
(i.e $dh=0=dh'$) to be equivalent if their difference is given by
$h-h'=d k$ where $k $ is a left invariant form. The cohomology is
then the set of equivalence classes with this equivalence relation.
A particular example of a  non-trivial element of this cohomology
would be  a closed  $p+2$ left invariant form, $h_{p+2}$, which can
therefore be constructed from the left invariant Cartan  one forms,
such that $h_{p+2}=d h_{p+1}$ where the $p+1$ form $ h_{p+1}$ is not
left invariant and so  it cannot be written in terms of left
invariant Cartan 1-forms.

In our case, we have a  $H$ invariant $p+2$ form $h$ given by
$$
h={{-i}\over{p!}} \left(e^{\underline a_1}\wedge \ldots \wedge
e^{\underline a_p}\right)\wedge \bar e^{\underline \alpha}\wedge
\bar  e^{\underline \beta}(\Gamma_{\underline a_1,...\underline
a_p}C^{-1})_{\underline \alpha\underline \beta} \;.  \eqno{(4.9)}$$
It is closed if
$$
(\Gamma^{\underline a_1}C^{-1})_{(\underline \alpha\underline \beta}
(\Gamma_{\underline a_1,...\underline a_p}C^{-1})_{\underline
\gamma\underline \delta)}=0, \eqno{(4.10)}$$ where indices
$(\underline \alpha\underline \beta\underline \gamma \underline
\delta)$ are totally symmetric. It holds in the dimensions in which
the p-branes exist [16], that is $p=1,2$ mod $4$  provided we take a
spinor of the appropriate character.

Using the relation between  the spinorial and the Lorentz
transformation
$$
{\tilde\Phi}\Gamma_{\underline a}{\tilde\Phi^{-1}}=
{(\Phi^{-1})_{\underline a}}^{\underline b} \Gamma_{\underline b}
\eqno{(4.11)}$$ one can show that  $h$ is actually independent of
${\Phi_{\underline b}}^{\underline a}$ and so ${\varphi_a}^{b'}$ and
is given by
$$
h=-{{i}\over{p!}} \left(\pi^{\underline a_1} \wedge ...\wedge \pi^
{\underline a_p}\right)\wedge \bar \pi^{\underline \alpha} \wedge
\bar \pi^{\underline \beta} (\Gamma_{\underline a_1,...\underline
a_p}C^{-1})_{\underline \alpha\underline \beta}  \;. \eqno{(4.12)}$$
Since the from $h$ is closed it can be written as $h=d{\cal L}^{WZ}$
where ${\cal L}^{WZ}$ is a quasi-invariant $p+1$ form, $\delta
{\cal L}^ {WZ}=d \lambda$. An explicit form of the WZ lagrangian is
[17]
$$
{\cal L}^{WZ}= {{-1}\over{(p+1)!}} \sum_{r=0}^p(-1)^r
\left(\matrix{p+1\cr r+1}\right)\pi^{\underline a_p}\wedge \ldots
\wedge \pi^{\underline a_{r+1}}\wedge K^{\underline a_r}\wedge
\ldots \wedge K^{\underline a_1}\wedge K_{\underline a_1,...
\underline a_p}, \eqno{(4.13)}$$ where  we have used $
(-1)^{{p(p+1)\over 2}}=-1$ and $$ K^{\underline
a}=i\bar\theta\Gamma^{\underline a}d\theta,\qquad K_{\underline
a_1,...\underline a_p}=i\bar\theta \Gamma_{\underline
a_1,...\underline a_p}d\theta. \eqno{(4.14)}$$ and
$\left(\matrix{p+1\cr r+1}\right)$
is the binomial coefficient.
This result can be shown
using $d \pi^{\underline a}=dK^{\underline a}$ and an identity
obtained from (4.10),
$$
dK^{\underline a_1} K_{\underline a_1,...\underline a_p}
+K^{\underline a_1}dK_{\underline a_1,...\underline a_p}=0 \;.
\eqno{(4.15)}$$

Since the Goldstone fields  ${\varphi_a}^{b'}$ appear only in the
first action $A^{NG}$,  its equation of motion is found following
similar steps to those explained in section two. The equation of
motion implies that ${\varphi_a}^{b'}= -(\pi^{-1})_b{}^{i} \pi_i{}^
{b'}$ which is equivalent to setting $e^{a'}=0$. Using this
expression we may eliminate ${\varphi_a}^{b'} $ from the action to
find that $A^{NG}$ is now given by $$
A^{NG}=-T\int\;d^{p+1}\xi\;\sqrt{-\det G_{ij}}
\eqno{(4.13)}$$ where $G_{ij}$ is the induced metric
$$
G_{ij}={{\pi_i}}{}^{\underline a}{{\pi_j}}{}^{\underline b}
\eta_{{\underline a}{\underline b}} \eqno{(4.14)}$$
and ${{\pi_i}}{}^{\underline a}$ is the supervielbein given by
(4.5).

The total action is $A=A^{NG}+bA^{WZ}$ where $b$ is a real constant
without no restrictions at classical level. It is well know that the
total action is invariant under kappa transformations if $b=\pm 1$.

\medskip
{\bf 5. Discussion}
\medskip

In this paper we have used the theory of non-linear realisations to
give new actions for the bosonic  and super p-branes. These actions
depend on all the Goldstone fields of the theory including those for
the broken Lorentz transformations. The latter can be eliminated
algebraically using their equations of motion to recover the usual
actions. These constructions underline  the importance of including
such Goldstone fields in addition to those associated with the
breaking of translations and supersymmetries. We have also shown
that when considering supersymmetric theories that arise from
spontaneous symmetry breaking one does not have to introduce
superfields. One of the aims of this paper was to provide a more
systematic account of the dynamics of  the simplest branes from the
view point of the theory of non-linear realisations. It has
been proposed [12] that  the branes in M theory should possess
$E_{11}$ symmetry and one may hope that the lessons learnt in this
paper may be of use in this much more difficult construction.
\par
It would be interesting to see if the actions given in this paper
which include the Goldstone bosons associated with  broken Lorentz
algebras can be extended to encompass theories where  more general
automorphisms are present as advocated in [11,12].
\par
Taking a group element that includes fields  associated to unbroken
translations  and other generators in $L$ implies the presence of
local symmetries in the action. We will show in a forthcoming paper,
that such symmetries, that is brane volume diffeomorphism and kappa
transformations  can also be deduced from the theory of non-linear
realisations.

\vskip 10mm

{\bf Acknowledgements} We acknowledge discussions with Andr\'es
Anabal\'on, Jaume Gomis, Norisuke Sakai, Toine Van Proeyen, Paul
Townsend, Jorge Zanelli. Joaquim Gomis and Peter West would like to
thank the CECS, Valdivia, Chile, where part of this work was carried
out,  for their support and hospitality.   We also thank Benasque Center
of Science for their hospitality.
This work is supported
in part by the European EC-RTN network MRTN-CT-2004-005104, MCYT FPA
2004-04582-C02-01, CIRIT GC 2005SGR-00564. PW is supported by a
PPARC senior fellowship PPA/Y/S/2002/001/44. This work was in
addition supported in part by the PPARC grant PPA/G/O/2000/00451 and
the EU Marie Curie research training work grant HPRN-CT-2000-00122.

\vskip 10mm

{\bf Appendix  }
\medskip

In this appendix we wish to compute the Cartan forms using some
explicit parameterisations of the Lorentz group. We choose the group
element for the bosonic p-brane to be of the form
$$
g=g_0\;U,\qquad g_0=e^{ix^{\underline a} P_{\underline a}},\qquad
U=e^{{i\phi_{a}{}^{b'}  J^{a}{}_{b'}}}. \eqno(A.1)$$ The Cartan
forms are  defined by
$$
{\cal V}=-ig^{-1}dg=U^{-1}(-ig_0^{-1}dg_0)U-i U^{-1}dU. \eqno(A.2)$$
where $-ig_0^{-1}dg_0= dx^{\underline a} P_{\underline a}$. Using
the commutator relations of the Poincare algebra we find that
$$
U^{-1} P_a U = {(\cosh V)_a}^bP_b +\phi_{a}{}^{c'} {({{\sinh \tilde
V}\over{\tilde V}})_{c'}}^{b'} P_{b'} ={\Phi_a}^{\underline b}
P_{\underline b},
$$
$$
U^{-1}P_{a'}U= - {\phi^{c}{}_{a'}}{({{\sinh V}\over{V}})_{c}}^{b}
P_b +{(\cosh \tilde V)_{a'}}^{b'}P_{b'} ={\Phi_{a'}}^{\underline
b}P_{\underline b} \eqno(A.3),
$$
where
$$
{(V^2)_a}^b=-\phi_{a}{}^{c'}\phi^{b}{}_{c'}
=-(\phi\phi^T)_{a}{}^{b},\qquad {(\tilde
V^2)_{a'}}^{b'}=-\phi^{c}{}_{a'}\phi_{c}{}^{b'}
=-(\phi^{T}\phi{)_{a'}}{}^{b'}  \eqno(A.4)$$
where
${{(\phi^T)}_{c'}}^b={\phi^b}_{c'}=\eta^{ba}{{\phi_a}^{d'}}\eta_{d'c'}$.
We also use
$$
{(V^2)_a}^b{\phi_b}^{c'}={{\phi}_{a}}^{b'}{(\tilde V^2)_{b'}}^{c'}.
\eqno(A.5)$$

Consequently we find that the   Lorentz transformation
$e^{i\phi_{c}{} ^{d'}  J^{c}{}_{d'}}$ has the vector representation
$\Phi_{\underline a}{}^{\underline b} $ which is given by
$$
{\Phi}=\pmatrix{{\Phi_a}^b&{\Phi_{a}}^{b'}\cr
{\Phi_{a'}}^b&{\Phi_{a'}}^{b'}} =\pmatrix{{(\cosh V)_a}^b&
\phi_{a}{}^{c'} {({{\sinh\tilde V}\over {\tilde V}})_{c'}}^{b'} \cr
- {\phi^{c}{}_{a'}}{({{\sinh V}\over{V}})_{c}}^{b} & {(\cosh \tilde
V)_{a'}}^{b'} }. \eqno(A.6)$$ In the section two, this same Lorentz
transformation was  presented in an alternative parametrisation
given in equation (2.6)
$$
\Phi= \left(\matrix{I_1& \varphi \cr -\varphi^T &
I_2\cr}\right)\left(\matrix{B_1& 0\cr 0 &
B_2\cr}\right)=\left(\matrix{B_1& \varphi B_2 \cr -\varphi^TB_1 &
B_2\cr}\right) \eqno(A.7)$$ where
$B_1=(I_1+\varphi\varphi^T)^{-{1\over 2}}$ and
$B_2=(I_2+\varphi^T\varphi)^{-{1\over 2}}$. These latter quantities
are related to the matrices $\phi$ by
$$
{(B_1)_a}^b={(\cosh V)_a}^b,\quad {(B_2)_{a'}}^{b'}={(\cosh \tilde
V)_ {a'}}^{b'}
$$
$$
{(\varphi)_{a}}^{b'}= {\phi_a}^{c'}{({{\sinh \tilde V}\over{\tilde
V\cosh \tilde V}})_{c'}}^ {b'}, \qquad
{(\varphi^T)_{a'}}^{b}={(\varphi)^{b}}_{a'}. \eqno(A.8)$$ The last
term of equation (A.2) can be written as
$$
-iU^{-1}  d  U=d {\phi_{a}}{}^{b'} \int_0^1 d\alpha \; e^{-i{\alpha
\phi_{c}{}^{d'}  J^{c}{}_{d'}}}  J^{a}{}_{b'} e^{i\alpha
{\phi_{c}{}^{d'}  J^{c}{}_{d'}}}\;
$$
$$
=d {\phi^{aa'}}\left({({{\sinh W}\over{W}})_{aa'}}^{bb'}J_{bb'}-
{({{\cosh W-1}\over{W^2}})_{aa'}}^{bb'}
(J_{bc}{\phi^c}_{b'}+J_{b'c'}{\phi_b}^{c'} )\right) \eqno(A.9)$$
where
$$
(W^2{)_{aa'}}^{bb'}={(V^2)_a}^b {\delta_{a'}}^{b'}+{\delta_a}^b
{(\tilde V^2)_{a'}}^{b'} +2{\phi_a}^{b'}{\phi^b}_{a'}. \eqno(A.10)$$

Using these equations the Cartan forms are found to be given by
$$
e^a= dx^{\underline b}{\Phi_{\underline b}}^a= dx^b{(\cosh V )_b}^a
-dx^{b'} {\phi^{c}{}_{b'}}{({{\sinh V}\over{V}})_ {c}}^{a},
$$
$$
e^{a'} =dx^{\underline b}{\Phi_{\underline b}}^{a'}=
dx^b\phi_{b}{}^{c'} {({{\sinh \tilde V}\over{\tilde V}} )_{c'}}^{a'}
+dx^{b'}{(\cosh\tilde V)_{b'}}^{a'},
$$
$$
w^{aa'} = d\phi^{bb'}({{\sinh W}\over{W}}{)_{bb'}}^{aa'},
$$
$$
w^{ab} = -{1\over2}d\phi^{cc'}{({{\cosh
W-1}\over{W^2}})_{cc'}}^{aa'} {\phi^{b}}_{a'}-(a\leftrightarrow b),
$$
$$
w^{a'b'} = -{1\over2}d\phi^{cc'}{({{\cosh
W-1}\over{W^2}})_{cc'}}^{aa'} {\phi_{a}}^{b'}-(a'\leftrightarrow b'),
\eqno(A.11)$$
where $(a\leftrightarrow b)$ is the term in which $a$ and $b$ are replaced
in the previous term.
\medskip

{\bf References}
\medskip
\noindent {  }
\item{[1]}
   S.~Weinberg,
   {\it ``Nonlinear Realizations Of Chiral Symmetry,''}
   Phys.\ Rev.\  {\bf 166} (1968) 1568; 
  J.~S.~Schwinger,
  {\it``Chiral Dynamics,''}
  Phys.\ Lett.\ B {\bf 24}(1967) 473;
  {\it``Partial Symmetry,''}
  Phys.\ Rev. \ Lett.  {\bf 18}(1967) 923; 
  D.~B.~Fairlie and K.~Yoshida,
  ``Chiral symmetry from divergence requirements,''
  Phys.\ Rev.\  {\bf 174} (1968) 1816.

\item {[2]}
   S.~R.~Coleman, J.~Wess and B.~Zumino,
   {\it ``Structure Of Phenomenological Lagrangians. 1,''}
   Phys.\ Rev.\  {\bf 177} (1969) 2239,
C.~G.~.~Callan, S.~R.~Coleman, J.~Wess and B.~Zumino,
   {\it ``Structure Of Phenomenological Lagrangians. 2,''}
   Phys.\ Rev.\  {\bf 177} (1969) 2247.

\item{[3]}
   A.~P.~Balachandran, A.~Stern and C.~G.~Trahern,
   {\it ``Nonlinear Models As Gauge Theories,''}
   Phys.\ Rev.\ D {\bf 19} (1979) 2416.

\item{[4]}
   M.~Bando, T.~Kugo, S.~Uehara, K.~Yamawaki and T.~Yanagida,
   {\it ``Is Rho Meson A Dynamical Gauge Boson Of Hidden Local
Symmetry?,''}
   Phys.\ Rev.\ Lett.\  {\bf 54} (1985) 1215.

\item{[5]} J. Hughes and J.  Polchinski, {\it Partially broken global
supersymmetry and the superstring}, Nucl Phys  B278 (1986) 147.

\item{[6]} J. Gauntlett, K. Itoh, and P. Townsend, {\it Superparticle
with extrinsic curvature}, Phys. Lett. B238, (1990) 65.

\item{[7]}   E.~A.~Ivanov and V.~I.~Ogievetsky,
    {\it ``The Inverse Higgs Phenomenon In Nonlinear Realizations,''}
    Teor.\ Mat.\ Fiz.\  {\bf 25} (1975) 164.

\item{[8]}  J.~P.~Gauntlett, J.~Gomis and P.~K.~Townsend,
    {\it ``Particle Actions As Wess-Zumino Terms For Space-Time (Super)
Symmetry
    Groups,''}
    Phys.\ Lett.\ B {\bf 249} (1990) 255,\
  J.~P.~Gauntlett and C.~F.~Yastremiz,
 {\it``Massive Superparticle in D = (2+1) and Sigma Model Solitons,''}
  Class.\ Quant.\ Grav.\  {\bf 7} (1990) 2089.

\item {[9]}  J. Bagger, A. Galperin,
{\it
   ``Matter couplings in partially broken extended supersymmetry,''}
Phys. Lett. {\bf B 336} (1994) 25; {\it
   ``A new Goldstone multiplet for partially broken supersymmetry,''}
Phys. Rev. {\bf D 55} (1997)
1091;
{\it ``The tensor Goldstone multiplet for partially broken  supersymmetry,''}
Phys. Lett. {\bf B 412} (1997) 296;
S. Bellucci, E. Ivanov, S. Krivonos,
{\it ``Partial breaking of N = 1 D = 10 supersymmetry,''}
Phys. Lett. {\bf B 460} (1999)
348; E. Ivanov, S. Krivonos,
{\it ``N = 1 D = 4 supermembrane in the coset approach,''}
Phys. Lett.
{\bf B 453} (1999) 237;
S. V. Ketov,
{\it ``Born-Infeld-Goldstone superfield actions for gauge-fixed
D-5 and D-3 branes in 6d,''}
{ Mod. Phys.
Lett.\/} {\bf A14} (1999) 501.

\item{[10]}
I.~A.~Bandos and A.~A.~Zheltukhin,
   ``Green-Schwarz superstrings in spinor moving frame formalism,''
  Phys. Lett. B {\bf 288}(1992) 77 ,
 I.~A.~Bandos, D.~P.~Sorokin, M.~Tonin, P.~Pasti and D.~V.~Volkov,
   ``Superstrings and supermembranes in the doubly supersymmetric
geometrical approach,''
  Nucl. Phys. B {\bf 446}, 79 (1995).

\item{[11]}   P. West,   {\it ``Automorphisms, non-linear realizations
and branes,''}
    JHEP {\bf 0002} (2000) 024.

\item{[12]}  P. West, {\it "$E_{11}$ origin of Brane charges and U-
duality multiplets"}.
  JHEP {\bf 0408} (2004) 052.

\item{[13]}
    J.~Brugues, T.~Curtright, J.~Gomis and L.~Mezincescu,
    {\it ``Non-relativistic strings and branes as non-linear
realizations of  Galilei
    groups,''}
    Phys.\ Lett.\ B {\bf 594} (2004) 227.

\item{[14]}
    J.~Brugues, J.~Gomis and K.~Kamimura,
    {\it ``Newton-Hooke algebras, non-relativistic branes and
generalized pp-wave
    metrics,''}
    Phys.\ Rev.\ D {\bf 73} (2006) 085011.

\item{[15]}
   J.~A.~De Azcarraga and P.~K.~Townsend,
   {\it ``Superspace Geometry And Classification Of Supersymmetric
Extended
    Objects,''}
    Phys.\ Rev.\ Lett.\  {\bf 62} (1989) 2579.

\item{[16]}
    A.~Achucarro, J.~M.~Evans, P.~K.~Townsend and D.~L.~Wiltshire,
    {\it ``Super P-Branes,''}  Phys.\ Lett.\ B {\bf 198} (1987) 441.

\item{[17]}
   J.~M.~Evans,
   {\it ``Super P - Brane Wess-Zumino Terms,''}
   Class.\ Quant.\ Grav.\  {\bf 5} (1988) L87.

\end